\documentclass[a4paper,11pt]{article}
\usepackage{pos}
\usepackage{hyperref}
\usepackage{soul}
\usepackage{subcaption}

\title{$T_{cc}^+$ via the plane wave approach and including diquark-antidiquark operators}

\author*[a,b]{Ivan Vujmilovic}
\author[c]{Sara Collins}
\author[a,b]{Luka Leskovec}
\author[a,e]{Emmanuel Ortiz-Pacheco}
\author[d]{M. Padmanath}
\author[a,b]{Sasa Prelovsek}

\affiliation[a]{Jozef Stefan Institute,\\
  Jamova cesta 39, 1000 Ljubljana, Slovenia}

\affiliation[b]{Faculty of Mathematics and Physics, University of Ljubljana,\\
Jadranska ulica 19, 1000 Ljubljana, Slovenia}

\affiliation[c]{Institut für Theoretische Physik, Universität Regensburg, 93040 Regensburg, Germany}

\affiliation[d]{The Institute of Mathematical Sciences, a CI of Homi Bhabha National Institute, Chennai, 600113, India}

\affiliation[e]{Department of Physics and Astronomy, Michigan State University, East Lansing, 48824, MI, USA}

\emailAdd{ivan.vujmilovic@ijs.si}
\emailAdd{sara.collins@ur.de}
\emailAdd{luka.leskovec@ijs.si}
\emailAdd{ortizpac@msu.edu}
\emailAdd{padmanath@imsc.res.in}
\emailAdd{sasa.prelovsek@ijs.si}

\abstract{The determination of the $DD^{*}$ scattering amplitude from lattice QCD is complicated by long-range interactions. In particular, the Lüscher method is no longer applicable in the kinematical region close to the left-hand cut. We tackle this problem by adopting  plane-wave and effective-field-theoretic methods, which also address partial wave mixing. In addition, we incorporate a diquark-antidiquark interpolator in the operator basis (along with the relevant scattering operators) in order to achieve a better resolution of the energy spectrum. Results show that inclusion of it already has some impact at physical charm quark mass, although it is more significant for larger heavy quark masses, in line with expectations.}

\FullConference{%
  The 41st International Symposium on Lattice Field Theory (LATTICE2024)\\
  28 July - 3 August 2024\\
  Liverpool, UK
}

\begin{document}
\maketitle
\section{Introduction}
\label{sec:intro}
The doubly-charmed tetraquark $T_{cc}^{+} (3875)$ was observed by the LHCb collaboration \cite{LHCb:2021auc, LHCb:2021vvq}, one of many exotic hadrons that have been discovered recently. The tetraquark state is a resonance located below the $D^{*+}D^{0}$ threshold and above the $D^0D^0\pi^{+}$ threshold. It  has isospin $I=0$, according to LHCb data, while its spin and parity are theoretically expected to be $J^P=1^+$, although these have not been experimentally measured yet. $T_{cc}^{+}$ has been subject of various lattice QCD studies (see e.g. refs. \cite{Padmanath:2022cvl, Junnarkar:2018twb, Collins:2024sfi}, among others). Because the resonance's mass lies only $0.36\left( 4 \right) \ \mathrm{MeV}$ below the nearest $D^{*+}D^{0}$ threshold, it is suspected to possess the structure of a meson-meson molecule. However, diquark-antidiquark is another type of binding that is consistent with the quantum numbers and quark content of the state under consideration. We  include interpolators of this type in the operator basis and study their effect on the spectrum and the resulting amplitudes. \\
\indent All lattice computations are performed at larger-than-physical pion masses, where the $D^{*}$ meson is stable. One can show that this induces a left-hand cut in the partial-wave projected $DD^*$ scattering amplitude, beginning at energies just under the $DD^*$ threshold, which is a well-known consequence of one-pion exchange in the $u$ channel. This invalidates the application of the usual tools employed in the extraction of scattering amplitudes from finite-volume spectra, such as Lüscher's quantization condition \cite{Luscher:1986pf}. In this work, this issue is addressed by using an alternative formalism. We adopt an effective potential description of $DD^*$ scattering and solve the Lippmann–Schwinger equation: first in finite-volume, in order to fit the parameters of the potential to lattice data, and then finally in infinite volume to find the pole in the scattering amplitude.

\section{Lattice Setup}
The numerical simulations were performed on two $N_f = 2  +  1$ CLS gauge field ensembles labelled U101 and H105. They share the same pion mass $m_{\pi} = 280(3) \ \mathrm{MeV}$ and lattice spacing $a = 0.08636(98)(40) \ \mathrm{fm}$ but have different spatial extents with $N_{L} = 24$ and $32$, respectively. Preliminary results of the finite-volume spectrum with diquark-antidiquark interpolators are also presented in \cite{Ortiz-Pacheco:2023ble}. In the following two sections, we give details of our operator basis and the extraction of the finite volume spectrum, which we then employ in our EFT plane-wave approach.

\section{Operator basis}
The basis employed in determining the finite-volume spectrum consists of operators that broadly fit into two distinct categories: bilocal meson-meson interpolators and one local diquark-antidiquark interpolator. While the physical significance of the former in studying the $T_{cc}^{+}$ tetraquark state is clear, the role of diquark-antidiquark interpolators, at least in terms of reliably determining the finite volume spectrum, is not completely resolved. 
Most lattice studies of $T_{bb}$ \cite{Leskovec:2019ioa, Alexandrou:2024iwi}, however, find that they are essential. \\
\label{subsubsec:MM}
\indent Bilocal meson-meson intepolators with the quantum numbers of $T_{cc}^+$ follow the general form
\begin{align*}\tag{1}
    \mathcal{O}^{MM}_{I = 0} \left( \vec{p}_1, \vec{p_2} \right) = \sum_{\vec{x}_1} e^{i \vec{p}_1 \cdot \vec{x}_1} \bar{u}(x_1) \Gamma_1 c(x_1) \sum_{\vec{x}_2} e^{i \vec{p}_2 \cdot \vec{x}_2} \bar{d}(x_2) \Gamma_2 c(x_2) - \{ u \leftrightarrow d \},
    \label{eq:MMop}
\end{align*}
where one selects various gamma matrices $\Gamma_{1, 2}$ and combinations of total momentum $\vec{P} = \vec{p}_1 + \vec{p}_2$ in order to couple to different partial waves $l = 0, 1, 2$ and parities. Our basis of scattering interpolators is largely unchanged with respect to the one used in refs. \cite{Padmanath:2022cvl, Ortiz-Pacheco:2023ble}, with a $D^{*}D^{*}$ interpolator projected to nonzero momentum $|\vec{P}| = 2 \pi / L$ being the only new addition. We apply ‘distillation’ smearing \cite{HadronSpectrum:2009krc} to all quark fields with the number of Laplacian eigenvectors used equal to 60 (90) on the ensemble with $N_L = 24$ ($32$). \\ 
\label{subsubsec:DDdd}
\indent In addition to scattering operators, we also employ a local diquark-antidiquark interpolator
\begin{align*}\tag{2}
    \mathcal{O}^{4q}_{I = 0} \left( \vec{p} \right) = \sum_{\vec{x}} \epsilon^{abc}\epsilon^{ade} \left[ c^{b}_{\alpha} (\vec{x}) (C \gamma_i)_{\alpha \beta} c^{c}_{\beta} (\vec{x}) \right] \left[ \bar{u}^{d}_{\delta} (C\gamma_5)_{\delta \sigma} \bar{d}^{e}_{\sigma} \right] e^{i \vec{p} \cdot \vec{x}}, 
    \label{eq:DDddop}
\end{align*}
where $C$ is the charge conjugation matrix. The number of Laplacian eigenvectors  is equal to 45 (55) on $N_L = 24$ ($32$). Even with a smaller number of eigenvectors, the computational cost of calculating correlators that contain the diquark-antidiquark operator is significantly higher compared to that for meson-meson interpolators. 

\section{Finite-volume energy spectrum}
Finite volume energies $E_n^{FV}$ and overlap factors $Z^{i}_{n}$  are extracted using the decomposition of two-point correlators into the eigenstates of the QCD Hamiltonian
\begin{align*}\tag{3}   
    C_2^{ij} (t) = \langle \Omega | \mathcal{O}^i (t) \mathcal{O}^{j\dagger} (0) | \Omega \rangle = \sum_{n} Z^{i}_{n} Z^{j*}_{n} e^{-E_{n}^{FV} t},
    \label{eq:2ptcorr}
\end{align*}
where $\mathcal{O}^i$ ($\mathcal{O}^{j\dagger}$) annihilate (create) the state with the quantum numbers of interest. The overlaps $Z^{i}_{n} = \langle \Omega | \mathcal{O}^{i} | n \rangle$ give the coupling of operators to the eigenstates. The energies and overlaps are extracted by solving the generalized eigenvalue problem (GEVP). 
\begin{figure}[h]
    \centering
    \includegraphics[width=1.0\textwidth]{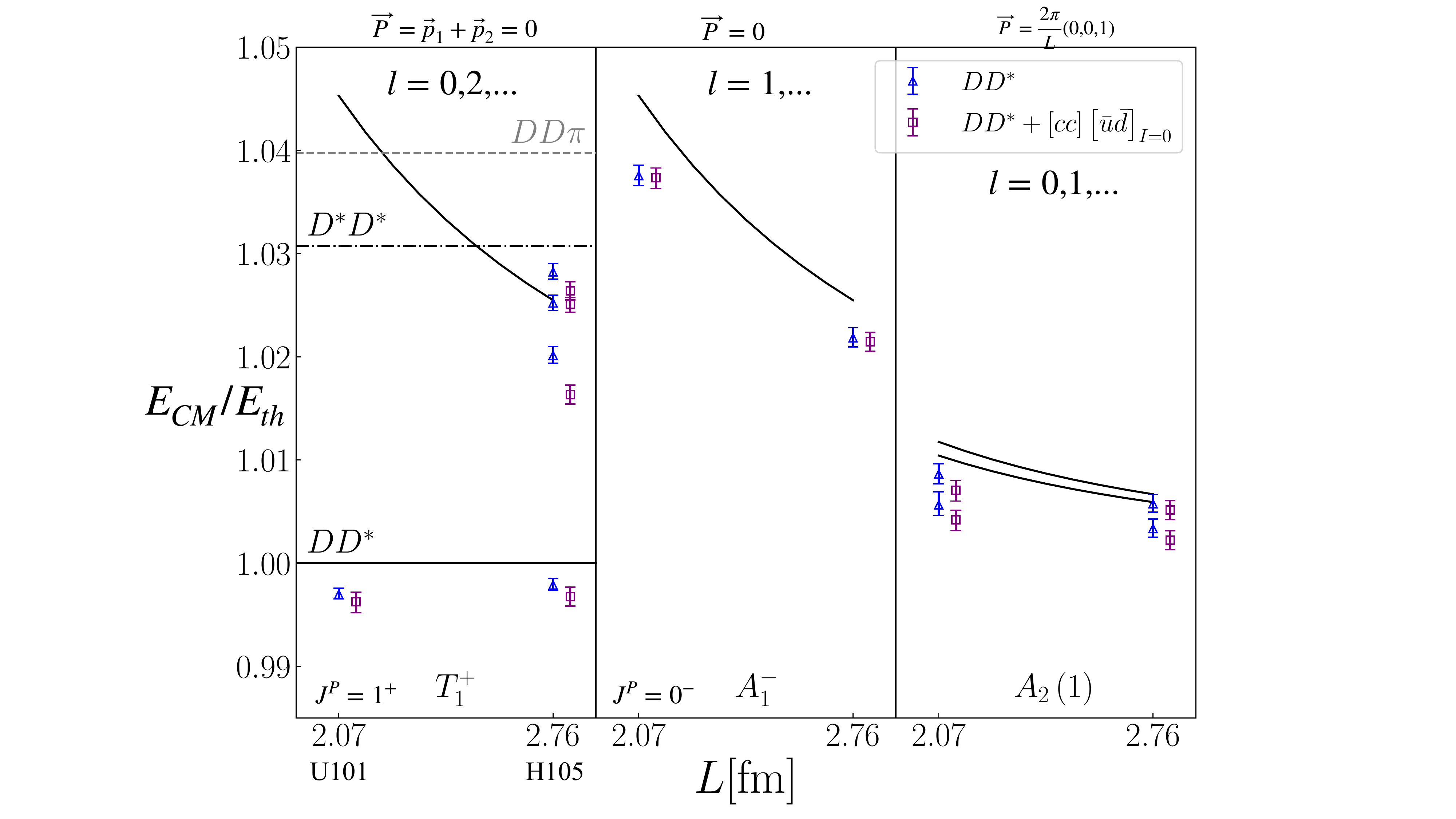}
    \caption{Finite-volume energy spectrum, shown for both ensembles. BLUE: energies in the center-of-momentum frame (CMF) obtained without the diquark-antidiquark interpolator~(eq. \eqref{eq:DDddop}) included in the operator basis. PURPLE: CMF energies obtained with the diquark-antidiquark interpolator included. Results are grouped into three different irreducible representations of the octahedral group ($T_1^+$, $A_1^-$) or its subgroup ($A_2 (1)$). Relevant $DD^{*}$, $D^{*}D^{*}$ and $DD\pi$ thresholds are also shown together with full lines indicating noninteracing levels.}
    \label{fig:FVenergyspec}
\end{figure}
The resulting energy spectrum is shown in Figure \ref{fig:FVenergyspec}. With the inclusion of the diquark-antidiquark interpolator we observe that the central values of almost all relevant energy levels remain unchanged within the 1$\sigma$ uncertainty. The only exception occurs for the second level in the $T_1^{+}$ irreducible representation of the larger ensemble with $N_L = 32$. We note that when employing only meson-meson interpolators in the analysis this level couples predominantly to the $D(1)D^{*}(-1)$ operator projected to $s$-wave, while the inclusion of the diquark-antidiquark interpolator introduces a strong coupling of the level to this interpolator and a small contribution from $D^{*}(0)D^{*}(0)$. Since we consider single-channel $DD^{*}$ scattering, we only make use of low-lying energy states that are significantly below the $D^{*}D^{*}$ threshold and do not couple significantly to the $D^{*}D^{*}$ interpolators. \\
\indent The same picture does not appear to hold when the heavy quark mass is increased to match the mass of the $b$ quark. We observe several energy levels that couple to both $BB^{*}$ and $B^{*}B^{*}$ interpolators, as well as diquark-antidiquark. The $BB^*$ and $B^*B^*$ thresholds also lie much closer together which indicates that a coupled channel analysis  is necessary in this case. Furthermore, the ground state of the system is significantly affected by the inclusion of diquark-antidiquark interpolators. More details will be provided in a future publication.
\section{Effective field theory and plane-wave approach}
\label{sec:eftpart}
As described in Section~\ref{sec:intro}, the main obstacle to the applicability of Lüscher's formalism is the existence of a left-hand cut resulting from one-pion exchange (OPE) illustrated in the right plot of Figure \ref{fig:effpot}. An effective potential of this interaction can be derived from the Lagrangian \cite{Fleming:2007rp}
\begin{align*}\tag{4}
    \mathcal{L}_{int} = \frac{g}{2 f_{\pi}} \left( D^{* \dagger} \cdot \nabla \pi^{a} \tau^{a} D + \mathrm{h. \ c.} \right), \ \mathrm{with} \ \pi^{a} \tau^{a} = \begin{pmatrix}
\pi^{0} & \sqrt{2} \pi^{+} \\
\sqrt{2} \pi^{-} & - \pi^{0}
\end{pmatrix}, 
    \label{eq:lagr}
\end{align*}
where $f_{\pi} = 92.2 \ \mathrm{MeV}$ and $g = 0.645$, whose value is based on \cite{Becirevic:2012pf}. The potential between $DD^{*}$ mesons is then given as
\begin{align*}\tag{5}
    V_{\pi} \left( \vec{q} = \vec{p}_2 - \vec{p}_1; \vec{\epsilon}_1, \vec{\epsilon}_2 \right) = 3 \left( \frac{g}{2 f_{\pi}} \right)^{2} \frac{\left( \vec{\epsilon}_1 \cdot \vec{q} \right) \left( \vec{\epsilon}_{2}^{*} \cdot \vec{q} \right)}{u - m_{\pi}^2} .
    \label{eq:pot1}
\end{align*}
$S$-wave projection of the OPE potential, derived in \cite{Collins:2024sfi} features a left-hand cut beginning approximately at $p_{lhc}^{2} \approx - \frac{1}{4} \left( m_{\pi}^2 - (m_{D^{*}} - m_{D})^{2} \right)$, the point at which  the exchanged pion can come on shell. The cut then extends to $- \infty$ along the real line. On the two ensembles that we use, the meson masses are $m_{D} = 1.927(2) \ \mathrm{MeV}$ and $m_{D^{*}} = 2.049(2) \ \mathrm{MeV}$. This leads to the left-hand cut appearing approximately at $99.6\%$ of the threshold energy $m_{D} + m_{D^{*}}$, indicating that Lüscher's formalism could fail for the two subthreshold levels shown in the leftmost plot of Fig.~\ref{fig:FVenergyspec}. 
\subsection{Lippmann-Schwinger equation}
\label{sec:LSEsec}
\indent To circumvent the problem discussed above, we make use of the Lippmann-Schwinger equation (LSE) illustrated in the left plot of Figure \ref{fig:effpot}
\begin{align*}\tag{6}
    \hat{T} = \hat{V} + \hat{V} \hat{\mathcal{G}} \hat{T},
    \label{eq:LSE}
\end{align*}
where $\hat{T}$, $\hat{V}$ and $\hat{\mathcal{G}}$ denote the scattering amplitude, effective potential and $2$-body propagator, respectively. In the nonrelativistic limit, the propagator  reduces to the Green's function of Schrödinger's equation
\begin{align*}\tag{7}
    \mathcal{G} (p^0, \vec{p}) = \frac{1}{ p^0  - \frac{\vec{p}^2}{2 m_r} + i \epsilon}, 
    \label{eq:prop}
\end{align*}
where $m_r = (\frac{1}{m_1} + \frac{1}{m_2})^{-1}$ is the  reduced mass of the system. \\
\indent Poles of the scattering amplitude $\hat{T}$ are determined by eq. \eqref{eq:LSE} as $\hat{T} = \hat{\mathcal{G}}^{-1} \left( \hat{\mathcal{G}}^{-1} - \hat{V} \right)^{-1} \hat{V}  \rightarrow \mathrm{det}\left( \hat{\mathcal{G}}^{-1} - \hat{V} \right) = 0$,
which in turn leads to the familiar Hamiltonian equation when the propagator in eq. \eqref{eq:prop} is inserted into the determinant equation. Upon defining $\hat{H} = \frac{\hat{p}^2}{2 m_r} + \hat{V}$ we arrive at
\begin{align*}\tag{8}
    \mathrm{det} \left( \hat{H} - p^0 \hat{I} \right) = 0.
    \label{eq:hameq}
\end{align*}
The same relation also holds in finite-volume when it is projected to appropriate irreducible representations of the octahedral group $O_h$ or one of its little groups 
\begin{align*}\tag{9}
    \mathrm{det} \left( \hat{H}^{\Gamma} - p^{0, \Gamma} \hat{I} \right) = 0 .
    \label{eq:hameqirr}
\end{align*}
Eq.~\eqref{eq:hameqirr} is fulfilled precisely when $p^{0, \Gamma}$ is equal to one of the energies of the finite-volume levels $E^{FV}_n$. 
\subsection{Effective potential}
\label{subsec:effpot}
\begin{figure}[h]
    \centering
    \begin{subfigure}[b]{0.54\textwidth}
        \centering
        \includegraphics[width=\textwidth]{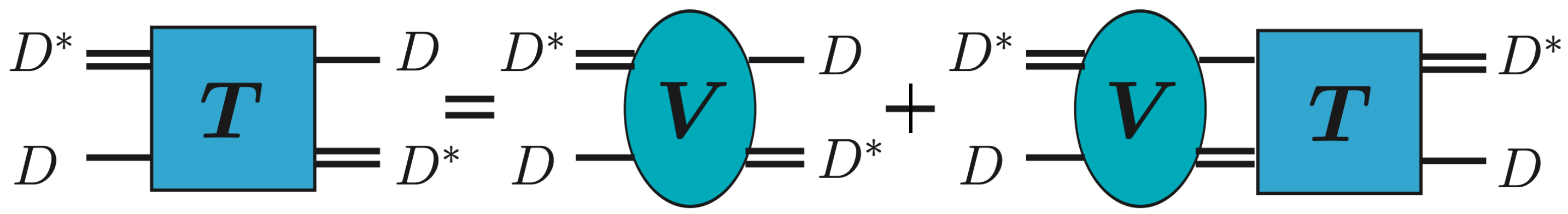}
    \end{subfigure}
    \begin{subfigure}[b]{0.45\textwidth}
        \centering
        \includegraphics[width=\textwidth]{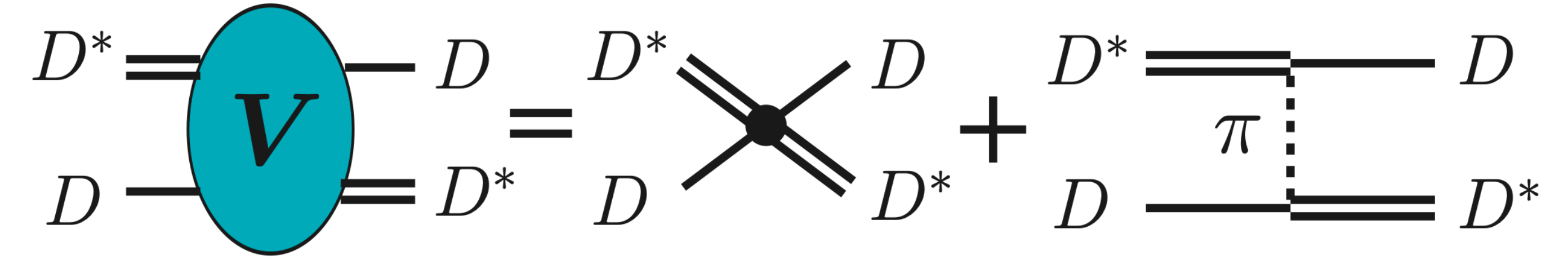}
    \end{subfigure}
    \caption{LEFT: Lippmann-Schwinger relation between $T$ and $V$. RIGHT: Effective potential $V$ defined in eq.~\eqref{eq:effpot}.}
    \label{fig:effpot}
\end{figure}

\indent The application of the Lippmann-Schwinger equation necessitates the use of an effective potential in parametrizing the interaction present in the $DD^{*}$ system. A suitable potential that describes finite-volume data well and explicitly incorporates one-pion exchange can be obtained from a low-energy expansion within the framework of chiral EFT (see ref. \cite{Meng:2023bmz} for more details)
\begin{align*}\tag{10}
    V \left( \vec{p}_1, \vec{\epsilon}_1; \vec{p}_2, \vec{\epsilon}_2 \right)
    \label{eq:effpot} = &\left( 2 c_0^s + 2 c_2^s (\vec{p}_1^2 + \vec{p}_2^2 ) \right) \left( \vec{\epsilon}_1 \cdot \vec{\epsilon}^{*}_2 \right) +  
     2 c_2^p  \left( \vec{p}_2 \cdot \vec{\epsilon}_{2}^{*} \right) \left( \vec{p}_1 \cdot \vec{\epsilon}_{1} \right) + \\[4pt]  
    &+3 \left( \frac{g}{2 f_{\pi}} \right)^{2} \frac{\left( \vec{\epsilon}_1 \cdot \vec{q} \right) \left( \vec{\epsilon}_{2}^{*} \cdot \vec{q} \right)}{u - m_{\pi}^2} ,
    \label{eq:fullpot}
\end{align*}
which is also shown illustrated on the right plot of Fig.~\ref{fig:effpot}. \\
\indent With the inclusion of the term in the second line of \eqref{eq:effpot} we explicitly account for the existence of the left-hand cut and incorporate it in our search for the pole in the $DD^{*}$ scattering amplitude; the $DD^{*}\pi$ coupling that appears in the OPE term is held fixed throughout all fits, $g = 0.645$ \cite{Becirevic:2012pf}. Note that the behavior of the one-pion exchange potential changes significantly at the unphysical pion masses of our simulation, $m_{\pi}^{lat} \approx 280$ MeV, compared to the physical case. In our setup it contributes a slight repulsion at long range and attraction at short range, while in the physical case it acts as a purely attractive force. Additionally, three low-energy constants (LECs) are introduced: the first term in the first line, described by two parameters $c_0^s$ and $c_2^s$, contributes to interaction in the $s$-wave and second term in the first line couples to the $p$-wave.
\subsection{Plane wave basis}
The natural choice of basis in which to evaluate matrices that appear in the Lippmann-Schwinger equation is composed of plane waves defined as
\begin{align*}\tag{11}
    | \vec{p}_1; \  \vec{p}_2, i \rangle, \ \mathrm{with} \ \vec{p}_i = \frac{2\pi}{L}\vec{n}_i, \ \vec{n}_i \in Z^3 \ \mathrm{and}  \ i = x, y, z,
\end{align*}
where $\vec{p}_1$ denotes the $3$-momentum of the $D$ meson and $\vec{p}_2, \ i$ denote the $3$-momentum and polarization of the $D^{*}$ meson, respectively \cite{Meng:2021uhz}. 
The use of plane-waves inherently implies the existence of a cutoff in this setup because of the necessary step of excluding momentum shells for which $||\vec{p}_i|| > p_{max}$ holds. In our analysis, choosing a cutoff that stays close to the $D^{*}D^{*}$ threshold and keeps everything nonrelativistic makes the most sense from a practical standpoint. Equivalently, we employ a sharp regulator of the form $F(p_1, p_2; \Lambda) = e^{ -\frac{p_1^n + p_2^n}{\Lambda^n}}$ with $n = 40$ and $\Lambda = 0.65 \ \mathrm{GeV}$ which multiplies the entire potential \eqref{eq:effpot}, i.e. $\tilde{V} = V \cdot F(p_1, p_2; \Lambda)$. 
The relevant cubic symmetry group is consequently dictated by the total momentum $\vec{P}$. The projection technique used to arrive at eq.~\eqref{eq:hameqirr} for various irreducible representations is well-established and explained in detail in refs. \cite{Meng:2021uhz, Prelovsek:2016iyo}.

\section{Results}

\begin{figure}[h]
    \centering
    \includegraphics[width=0.95\textwidth]{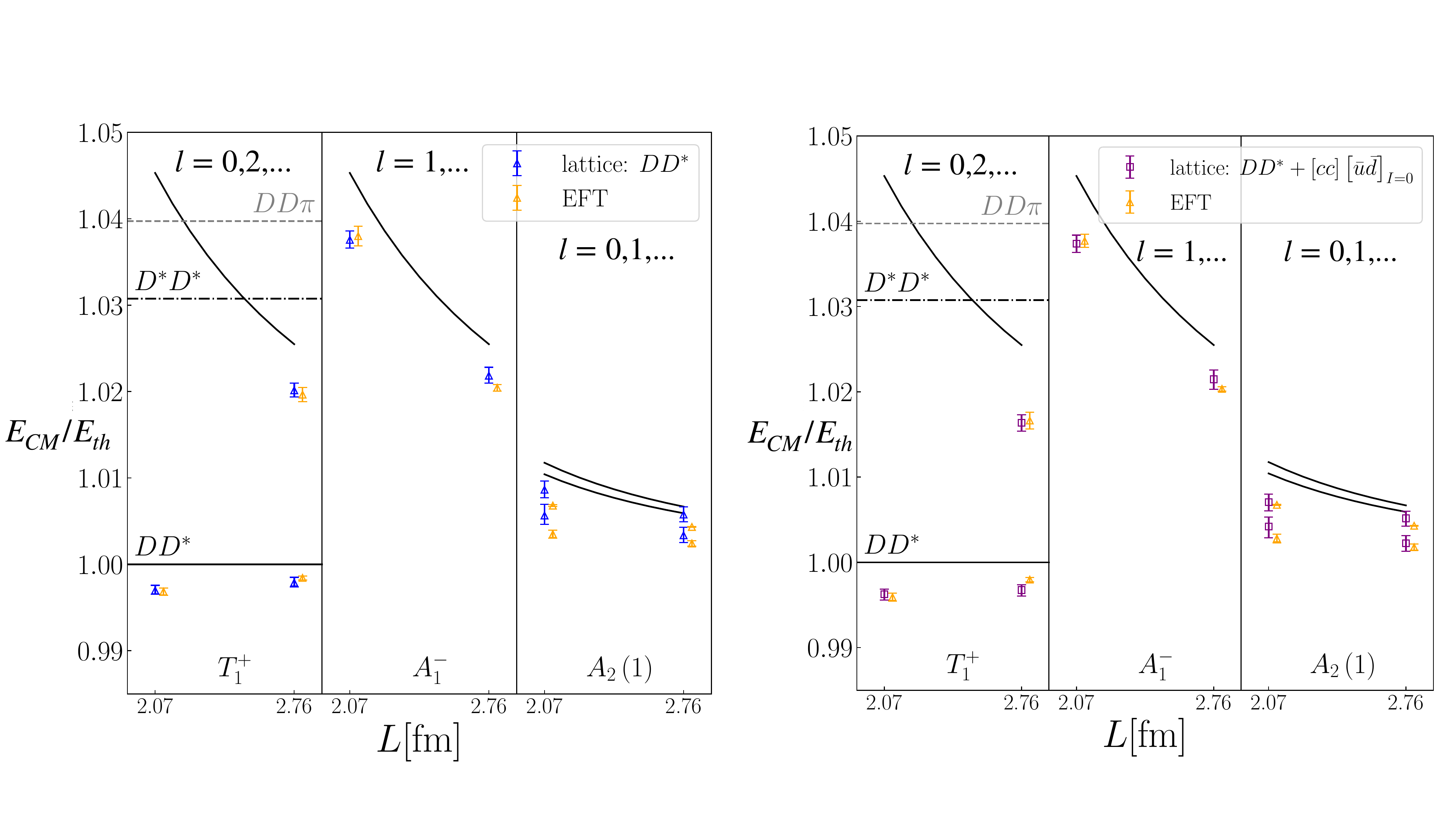}
    \caption{Fit results for the energies in finite-volume. Blue and purple points represent energies obtained from the correlators, while orange points are the results calculated from the effective potential and eq.~\eqref{eq:hameqirr}. Left plot shows the best fit to only meson-meson data and right plot includes the diquark-antidiquark interpolator.}
    \label{fig:lecfit}
\end{figure}

In the first step, the finite volume Lippmann-Schwinger equation, reduced to a Hamiltonian equation \eqref{eq:hameqirr}, is solved in the first step.
This allows us to fit the three low-energy constants of the effective potential to lattice data. Both fits are displayed in Figure \ref{fig:lecfit}. While the model from Section~\ref{sec:eftpart} describes meson-meson data reasonably well, the fit is significantly improved with the inclusion of diquark-antidiquark operator, lowering the reduced $\chi^2 / n_{dof}$ from $2.4$ to $1.36$.

\subsection{Comparison of EFT plane-wave approach and Lüscher's formalism}
\begin{figure}[h]
    \centering
    \includegraphics[width=0.95\textwidth]{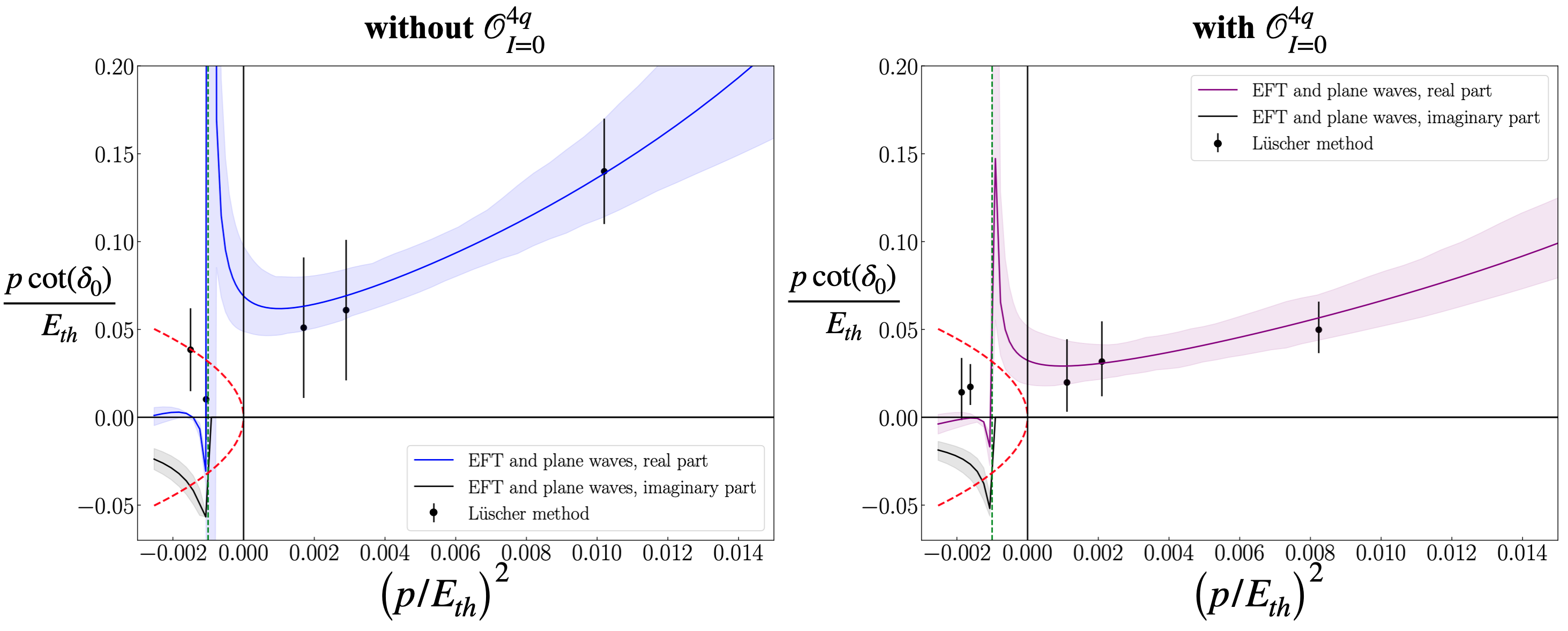}
    \caption{Comparison of $s$-wave phase shifts. The left plot features only meson-meson data and the right plot includes the diquark-antidiquark interpolator. The vertical green line marks the beginning of the left-hand cut and the red line indicates $i p / E_{th}$.}
    \label{fig:eftluscher}
\end{figure}

The comparison of $s$-wave phase shifts $p\cot ( \delta_0 ) / E_{th}$ generated by the plane-wave EFT approach and Lüscher's formula is shown in Figure \ref{fig:eftluscher}. Note that, in line with expectations, both methods are in good agreement at energies above the left-hand cut, marked with a vertical green line on both plots.

\subsection{$T_{cc}^{+}$ pole}

\begin{figure}[h]
    \centering
    \includegraphics[width=0.80\textwidth]{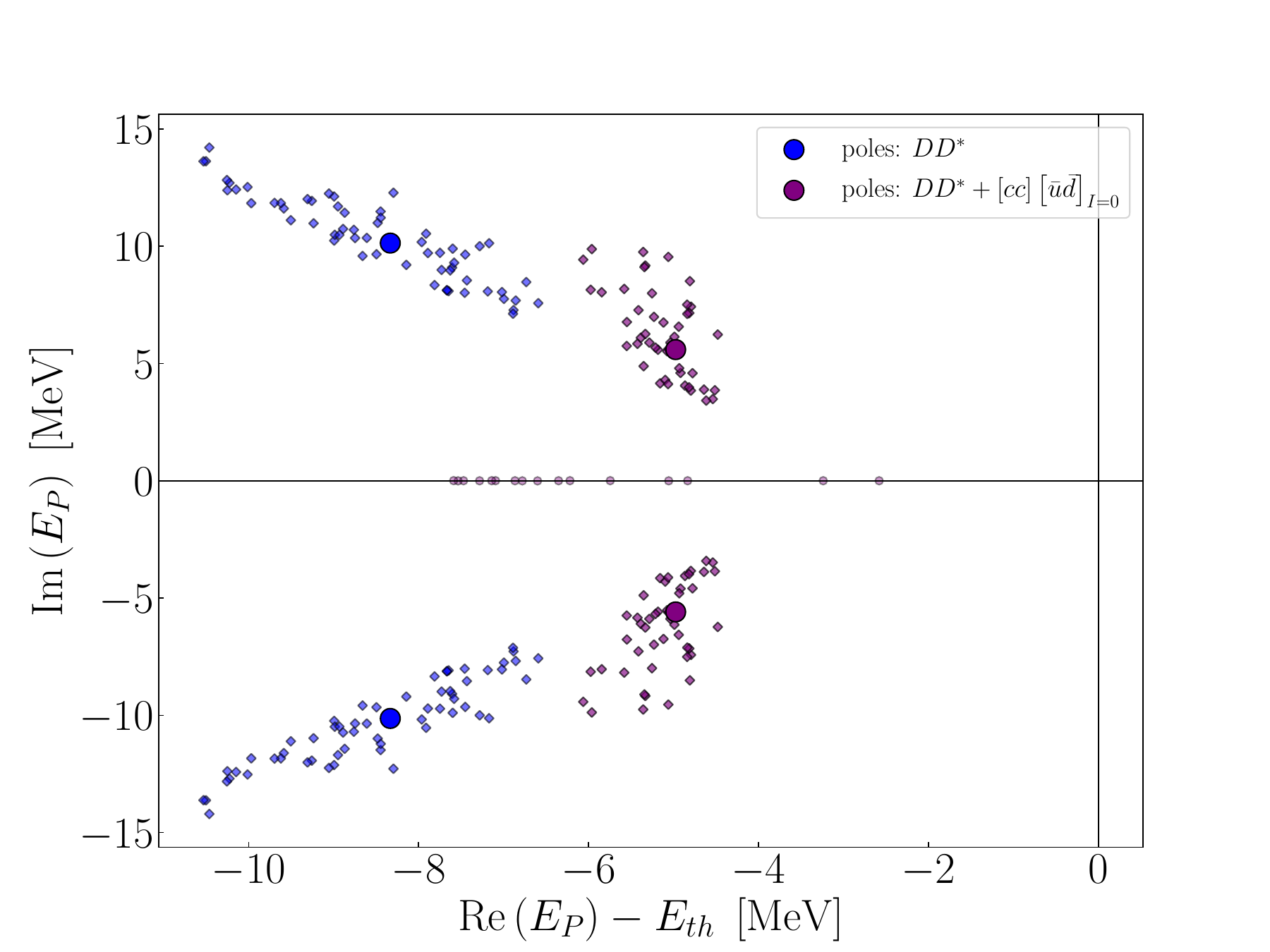}
    \caption{Location of the $T_{cc}^{+}$ pole in the infinite volume continuum. Central values are represented by two pairs of bigger blue/purple dots and smaller diamonds are obtained by varying the LECs within their error bounds. The origin represents $DD^{*}$ threshold.}
    \label{fig:tccpole}
\end{figure}

The final result on the location of the $T_{cc}^{+}$ pole is presented in Figure \ref{fig:tccpole}. In both sets of data the tetraquark appears as a subthreshold resonance, which is to a large extent explained by the interplay of the one-pion exchange potential \eqref{eq:pot1} and the system's kinematics, as discussed in \cite{Collins:2024sfi}. Moreover, the data with the diquark-antidiquark interpolator included shows greater attraction present in the system, shifting the pole location closer towards the threshold and the physical scattering axis, which shows that it has a noticeable impact in describing the $T_{cc}^{+}$.

\section{Conclusion and outlook}
In this work we have presented results of our study of the $T_{cc}^{+}$ tetraquark. Building upon the already existing operator basis of scattering interpolators from ref. \cite{Padmanath:2022cvl}, we implemented an additional local diquark-antidiquark interpolator. The lattice data was analyzed in a framework combining effective field theory and plane-waves to find the position of the $T_{cc}^{+}$ resonance in the $DD^{*}$ scattering amplitude. We find that the diquark-antidiquark operator has a somewhat small,
non-negligible impact on the pole, with the system exhibiting greater attraction with its inclusion. On the other hand this effect is more pronounced when the heavy quark mass is closer to the $b$ quark mass, where we observe a significant shift in the ground state energy.

\section*{Acknowledgments}
We would like to thank V. Baru, S. Dawid, E. Epelbaum, L. Meng, A. Nefediev, F. Romero-López and S. Sharpe for illuminating discussions. The authors gratefully acknowledge the HPC RIVR consortium (\href{https://www.hpc-rivr.si}{www.hpc-rivr.si}) and EuroHPC JU (\href{https://eurohpc-ju.europa.eu/}{eurohpc-ju.europa.eu}) for funding this research by providing computing resources of the HPC system Vega at the Institute of Information Science (\href{www.izum.si}{www.izum.si}). The work of I. V., S. P. and L. L. is supported by the Slovenian Research Agency (research core Funding No. P1-0035 and J1-3034 and N1-0360).

\end{document}